\documentclass{article}

\usepackage[a4paper,margin=1in]{geometry}
\usepackage{amsmath,amssymb,bm}
\usepackage{tikz}
\usetikzlibrary{shapes.geometric,positioning}

\newcommand{\supp}{\operatorname{supp}}

\usepackage{amsmath}
\usepackage{amssymb}
\usepackage{bm}

\usepackage{graphicx}
\usepackage{booktabs}
\usepackage{siunitx}

\usepackage[font=small]{caption}
\usepackage{subcaption}

\usepackage[ruled,vlined,linesnumbered]{algorithm2e}

\usepackage{authblk}

\usepackage{hyperref}

\title{Performance of Planar XYZ and XZZX Codes under Homogeneous and Spatially Nonuniform Noise}
\author[1]{Zhiwei Wang}
\author[1]{Liqi Wang}
\affil[1]{School of Mathematics, Hefei University of Technology, 230601 Hefei, China}

\begin{document}
\maketitle

\begin{abstract}
The minimum-weight perfect matching (MWPM) decoder is a standard decoding strategy 
for surface codes, but its performance can degrade considerably under biased noise. 
In this work, a modified family of surface codes, termed the XYZ planar code, is introduced, 
and MWPM is extended to posterior MWPM (pMWPM), which exploits the information provided by 
mixed stabilizers with almost no increase in decoding complexity. The XYZ--pMWPM and XZZX 
code--decoder combinations are compared under homogeneous independent and identically 
distributed (IID) Pauli noise and processor-derived independent nonidentically 
distributed (INID) noise. Under homogeneous biased Pauli noise, pMWPM substantially 
increases the threshold of the XYZ planar code relative to conventional MWPM. 
At finite code distances, particularly for small codes under strong bias, 
XYZ--pMWPM achieves logical error rates comparable to or lower than those of XZZX. 
Under INID noise, the pseudothreshold ordering observed under the corresponding homogeneous 
IID references is not always preserved; in some hardware--distance settings, 
XYZ--pMWPM achieves pseudothresholds comparable to or higher than those of XZZX. 
Within XZZX, qubit-specific weighting consistently yields higher INID pseudothresholds 
than average-rate weighting for the hardware instances considered. It is further found that 
the IID--INID pseudothreshold shift depends strongly on the physical-error scalarization and 
the construction of the homogeneous IID reference. These results show that pMWPM efficiently 
exploits the mixed-stabilizer structure of the XYZ planar code and provides competitive 
finite-distance decoding under biased and spatially nonuniform noise.
\end{abstract}

\section{INTRODUCTION}

Quantum error correction (QEC) stands as a critical enabler for the realization of scalable, fault-tolerant quantum computation.
Among various QEC codes, the surface code~\cite{kitaev2003fault,bravyi1998quantum} has emerged as a leading candidate due to its high error threshold and compatibility with local interactions in two-dimensional architectures.
However, its performance is typically optimized for symmetric noise models, such as depolarizing noise, which may not accurately reflect the physical noise characteristics of many quantum computing platforms.
Notably, in systems such as superconducting qubits~\cite{aliferis2009fault}, quantum dots~\cite{shulman2012demonstration}, and trapped ions~\cite{nigg2014quantum}, noise can be biased, with dephasing errors occurring at a significantly higher rate than bit-flip errors.

Recent research indicates that tailoring surface code variants to biased noise can yield significant performance benefits,
where bias is defined as the ratio of the probability of a high-rate \(Z\) error to the sum of the probabilities of low-rate \(X\) and \(Y\) errors.
The XY surface code doubles the number of effective syndrome bits associated with dominant \(Z\) errors by measuring the product of \(Y\) operators around the surface rather than \(Z\) operators~\cite{tuckett2018ultrahigh,tuckett2019tailoring}.
The stabilizers of the XZZX surface code consist of products of two \(X\) and two \(Z\) operators on each face~\cite{bonilla2021xzzx}.
Both the XY and XZZX surface codes are Clifford-deformed surface codes~\cite{dua2024clifford} exhibiting remarkably high thresholds, approaching or even saturating the hashing bound across a wide range of biases.
Despite these advantages, these constructions employ fixed stabilizer patterns designed for prescribed bias structures and do not explicitly account for device-specific spatial variations in the noise. In realistic processors, 
where error rates and bias may vary across qubits, incorporating such local noise information into the decoder can therefore be important for maintaining decoding performance.
These considerations motivate the joint adaptation of the stabilizer structure and decoding metric rather than relying on a single code deformation for all noise conditions.

Surface codes are typically decoded using a variety of methods.
The minimum-weight perfect matching (MWPM) decoder~\cite{edmonds1965paths,fowler2015minimum} has been accelerated by modern implementations such as sparse blossom~\cite{higgott2022pymatching,higgott2023sparse} and fusion blossom~\cite{wu2023fusion}, achieving near-linear expected complexity.
The union-find (UF) decoder~\cite{delfosse2021almost} offers nearly linear worst-case complexity. Belief propagation (BP) can also be combined with both weighted UF and MWPM, leading to belief-find and belief-matching decoders, 
respectively, in which BP-updated beliefs are used to determine the edge weights of the subsequent decoding graph~\cite{higgott2023improved}.
The BPOSD decoder~\cite{panteleev2021degenerate,roffe2020decoding} integrates BP with ordered statistics decoding to mitigate the degeneracy problem, reaching thresholds comparable to MWPM.
The tensor network (TN) decoder~\cite{tuckett2019tailoring,bravyi2014efficient} attains high thresholds under depolarizing noise, but at a significantly higher computational cost.
Other decoding methods include cellular-automaton decoders~\cite{herold2015cellular,kubica2019cellular}, renormalization group decoders~\cite{duclos2010fast}, neural-network decoders~\cite{varsamopoulos2018decoding,chamberland2018deep}, and MaxSAT decoders~\cite{berent2023decoding}.
MWPM therefore remains a strong candidate for real-time decoding due to its competitive decoding performance and the near-linear average complexity enabled by modern implementations.

This paper proposes two XYZ planar code constructions, referred to as the original and reoriented geometries, together with an improved MWPM strategy tailored to biased and spatially non-identical noise.
The approach draws inspiration from existing surface code modifications and from the reweighting and recursive MWPM strategies developed in Refs.~\cite{iolius2022performance,iolius2023performance}.
The former showed that qubit-resolved reweighting and optimized qubit placement can improve surface code performance under independent nonidentically distributed (INID) noise, 
whereas the latter further incorporated noise bias, nonuniformity, and correlations between Pauli error components into a recursive MWPM procedure.
In contrast, while these approaches retain a fixed surface code stabilizer structure and primarily adapt the decoder weights or qubit arrangement, 
the present work introduces two modified XYZ planar code constructions together with posterior-informed MWPM weighting. 
Thus, both the stabilizer structure and the matching metric are adapted to the underlying noise characteristics. 
In particular, the stabilizer arrangement is modified according to the relevant noise-bias tendency, 
while the decoder weights are determined using the corresponding physical error probabilities and bias information.

The resulting code-decoder combination is numerically investigated under both symmetric and biased Pauli channels 
and compared with conventional MWPM and the XZZX surface code. In addition to threshold calculations under homogeneous noise, 
finite-distance pseudothresholds are evaluated under both IID and INID noise, with the latter incorporating 
processor-specific, qubit-resolved \(T_1\) and \(T_2\) data reported for superconducting quantum processors~\cite{iolius2022performance}.
These numerical results assess both the 
performance benefits of the proposed construction under controlled bias and its robustness in the presence of spatial 
nonuniformity and processor-dependent noise characteristics.

\section{PRELIMINARIES}
\label{sec:preliminaries}

\subsection*{A. Planar code}
\label{subsec:planar_preliminaries}

The surface code is a family of topological stabilizer codes defined on
a two-dimensional lattice with local check
operators~\cite{dennis2002topological,terhal2015quantum,fowler2012surface}.
In this work, the planar code is used as the parent code for
the XYZ constructions introduced later. For odd code distance $d$, it
contains
\begin{equation}
n=d^2+(d-1)^2=2d^2-2d+1
\label{eq:planar_n}
\end{equation}
data qubits and $n-1$ independent stabilizer generators, and hence
encodes one logical qubit.

The planar code is a CSS stabilizer code with stabilizer group
$\mathcal S$ generated by local $X$- and $Z$-type checks,
\begin{equation}
S_x^X=\prod_{i\in\operatorname{supp}(x)}X_i,
\qquad
S_z^Z=\prod_{i\in\operatorname{supp}(z)}Z_i .
\label{eq:planar_checks_general}
\end{equation}
where \(\text{supp}(x) = \{i : x_i = 1\}\) and \(\text{supp}(z) = \{i : z_i = 1\}\) 
denote the supports of the binary vectors \(x\) and \(z\), respectively.
The generators commute pairwise, and the codespace is their simultaneous
$(+1)$ eigenspace~\cite{higgott2022pymatching}. A Pauli error $E$ is
characterized by the syndrome
\begin{equation}
\sigma_j(E)=
\begin{cases}
0, & [E,S_j]=0,\\
1, & \{E,S_j\}=0,
\end{cases}
\label{eq:syndrome_definition}
\end{equation}
so that $X$-type checks detect the $Z$ component of an error,
whereas $Z$-type checks detect its $X$ component. 

Logical operators belong to
$N(\mathcal S)\setminus\mathcal S$,
where \(N(\mathcal{S})\) denotes the normalizer 
of the stabilizer group \(\mathcal{S}\) in the \(n\)-qubit Pauli group.
A logical pair
may be chosen as
\begin{equation}
\overline X_{\rm P}
=
\prod_{i\in\gamma_X}X_i,
\qquad
\overline Z_{\rm P}
=
\prod_{i\in\gamma_Z}Z_i ,
\label{eq:planar_logical_general}
\end{equation}
where $\gamma_X$ and $\gamma_Z$ are nontrivial paths connecting the
corresponding opposite boundaries. Their minimum-weight representatives
have weight $d$ and satisfy
\begin{equation}
\overline X_{\rm P}\overline Z_{\rm P}
=
-\overline Z_{\rm P}\overline X_{\rm P}.
\end{equation}

Given a measured syndrome, a decoder returns a recovery operator $R$.
Decoding succeeds when
\begin{equation}
RE\in\mathcal S,
\label{eq:decode_success}
\end{equation}
whereas a logical failure corresponds to
\begin{equation}
RE\in L\mathcal S,
\qquad
L\in N(\mathcal S)\setminus\mathcal S .
\label{eq:decode_failure}
\end{equation}
Hence the recovery need not reproduce the physical error itself, but
only its logical equivalence class.

\subsection*{B. Minimum-weight perfect matching decoder}
\label{subsec:mwpm_preliminaries}

Minimum-weight perfect matching (MWPM) is a standard decoder for
surface codes~\cite{edmonds1965paths,higgott2022pymatching}.
For the planar code, the $X$ and $Z$ error components can
be decoded on separate matching graphs. In the graph used to decode
the $Z$ component, vertices correspond to $X$-type checks and a single-qubit 
error with a \(Z\) component connects the checks whose syndromes 
it changes; boundary errors are represented by edges terminating at an
appropriate boundary. The analogous construction using $Z$-type
checks is used to decode the $X$ component.

A measured syndrome identifies the set of check nodes associated with nonzero syndrome outcomes. 
MWPM then finds a minimum-weight collection of error edges whose boundary is precisely this set of check nodes. 
For statistically independent error mechanisms, an
edge associated with an error probability $q_i$ is assigned the
log-likelihood weight
\begin{equation}
w_i
=
\log\frac{1-q_i}{q_i}.
\label{eq:mwpm_weight}
\end{equation}
Indeed, for a binary error vector $\bm e$,
\begin{equation}
P(E)
=
\prod_i
(1-q_i)^{1-e_i}q_i^{e_i},
\end{equation}
and hence
\begin{equation}
\log P(E)
=
\sum_i\log(1-q_i)
-
\sum_i w_i e_i .
\label{eq:mwpm_log_probability}
\end{equation}
Minimizing the total matching weight therefore selects the most probable individual 
error configuration consistent with the observed syndrome under the independent graphlike error model.

For the planar code with perfect syndrome measurements, MWPM exhibits
a threshold of approximately $10.3\%$ for a single independent
$X$- or $Z$-error component; the corresponding total physical error
threshold under the conventional depolarizing model is approximately
$15.5\%$~\cite{wang2010threshold}. More generally, decoder performance
depends on both the Pauli bias and the spatial distribution of the
physical error probabilities~\cite{iolius2024review}. This motivates
using weighted matching graphs rather than uniform geometric weights
when the underlying qubits are not identically distributed.

The complexity of MWPM arises from two steps: constructing the syndrome graph (using Dijkstra's algorithm) and finding a perfect matching (using the blossom algorithm). 
The Dijkstra step has complexity \(O(n^2 \log n)\), while the blossom step dominates with \(O(n^3 \log n)\). If equal weights are assumed, the graph can be built using Manhattan distance, but the overall runtime remains \(O(d^6 \log d)\)~\cite{higgott2022pymatching}, where \(n\) is the number of data qubits and \(d\) is the code distance.

\subsection*{C. Probability pseudothreshold}
\label{subsec:pseudothreshold}

Under an IID noise model,
the probability threshold $p_{\mathrm{th}}$ is the critical physical
error rate below which increasing the code distance suppresses the
logical error rate. Accordingly, logical error curves for different
distances typically intersect near a common critical point~\cite{nielsen2011quantum}. 

In realistic quantum processors, the relaxation and dephasing times
can vary substantially among data qubits. The resulting noise can
therefore be modeled as INID. In contrast to an IID channel, each data qubit $i$
has its own $T_{1,i}$ and $T_{2,i}$, and hence its own Pauli error
probabilities.

In this work, the decoherence of each data qubit is described by the
Pauli-twirled amplitude--phase damping channel. After an elapsed time
$t$, the Pauli error probabilities of qubit $i$ are~\cite{iolius2022performance,iolius2023performance}
\begin{align}
p_{X,i}(t)
&=
p_{Y,i}(t)
=
\frac14
\left(
1-e^{-t/T_{1,i}}
\right),
\label{eq:px}\\
p_{Z,i}(t)
&=
\frac14
\left(
1+e^{-t/T_{1,i}}
-2e^{-t/T_{2,i}}
\right),
\label{eq:pz}
\end{align}
and its total probability of a nontrivial Pauli error is
\begin{equation}
p_i(t)
=
p_{X,i}(t)+p_{Y,i}(t)+p_{Z,i}(t).
\label{eq:pi}
\end{equation}
The individual probabilities in
Eqs.~(\ref{eq:px})--(\ref{eq:pi}) are used in the actual INID
noise sampling and, when applicable, in the calculation of
qubit-dependent decoder weights.

Unlike an IID channel, an INID channel does not possess a
unique single-qubit physical error probability. A scalar physical
error rate must therefore be defined when reporting logical error
curves and probability pseudothresholds. 
Two such definitions characterize different aspects of the heterogeneous noise.

\paragraph{Mean-\(T\) definition.}

Previous studies of surface codes under INID hardware noise
defined the physical error reference through an IID channel
constructed from the average relaxation and dephasing times~\cite{iolius2022performance,iolius2023performance}. 
Following this
convention, let
\begin{equation}
\overline T_1
=
\frac{1}{n}\sum_{i=1}^{n}T_{1,i},
\qquad
\overline T_2
=
\frac{1}{n}\sum_{i=1}^{n}T_{2,i}.
\label{eq:mean_coherence_times}
\end{equation}
The corresponding mean-\(T\) physical error parameter is
\begin{equation}
p_{\mathrm{meanT}}(t)
=
\frac34
-\frac14e^{-t/\overline T_1}
-\frac12e^{-t/\overline T_2}.
\label{eq:p_meanT}
\end{equation}

This definition provides a common time-derived reference associated
with the average coherence properties of the selected hardware
qubits. In particular, the corresponding IID reference channel is
obtained by assigning the same
$(\overline T_1,\overline T_2)$ to every data qubit. It therefore
answers the question of how a heterogeneous device performs relative
to a homogeneous device having the same average coherence times.

Importantly, $p_{\mathrm{meanT}}$ is a nominal reference for the
INID channel rather than its actual mean error probability. Since
the mapping from $(T_1,T_2)$ to Pauli error probabilities is nonlinear,
in general
\begin{equation}
p_{\mathrm{meanT}}(t)
\neq
\frac{1}{n}\sum_{i=1}^{n}p_i(t).
\label{eq:meanT_not_pmean}
\end{equation}

\paragraph{\(P\)-mean definition.}

To complement the mean-T convention, a second physical error parameter is introduced, 
obtained by averaging the actual local Pauli error probabilities:
\begin{equation}
p_{\mathrm{Pmean}}(t)
=
\frac{1}{n}
\sum_{i=1}^{n}p_i(t)
=
\frac{1}{n}
\sum_{i=1}^{n}
\left[
p_{X,i}(t)+p_{Y,i}(t)+p_{Z,i}(t)
\right].
\label{eq:p_Pmean}
\end{equation}
Equivalently, defining
\begin{equation}
\overline p_{\alpha}(t)
=
\frac{1}{n}\sum_{i=1}^{n}p_{\alpha,i}(t),
\qquad
\alpha\in\{X,Y,Z\},
\end{equation}
gives
\begin{equation}
p_{\mathrm{Pmean}}(t)
=
\overline p_X(t)
+
\overline p_Y(t)
+
\overline p_Z(t).
\end{equation}

Unlike $p_{\mathrm{meanT}}$, this quantity is the actual mean
probability that a data qubit undergoes a nontrivial Pauli error.
Equivalently, for independent single-qubit errors,
$n\,p_{\mathrm{Pmean}}$ is the expected number of erroneous data
qubits in one noise realization. The matched IID reference for
this definition assigns
$(\overline p_X,\overline p_Y,\overline p_Z)$ uniformly to all data
qubits. Consequently, the IID and INID channels have the same
mean Pauli error probabilities and differ only in their spatial
distribution. The \(P\)-mean construction controls the mean Pauli error vector, 
thereby providing a reference for examining spatial error rate 
heterogeneity at fixed mean Pauli error strength.

These two definitions are complementary rather than interchangeable.
The mean-\(T\) convention compares heterogeneous and homogeneous hardware at fixed average coherence parameters and preserves the reference used in previous work, 
whereas the \(P\)-mean convention compares them at fixed average Pauli error probability.
Their difference arises directly from the nonlinear transformation between
coherence times and error probabilities,
\begin{equation}
\frac{1}{n}\sum_i e^{-t/T_i}
\neq
e^{-t/\overline T},
\end{equation}
so the two scalar descriptions generally need not produce the same
pseudothreshold.

Because distance-dependent logical error curves under INID noise need not exhibit a unique common intersection, 
the conventional probability threshold does not provide an unambiguous finite-device reference~\cite{iolius2022performance}. 
Therefore, the probability pseudothreshold is used, 
defined as the point at which the logical error rate equals the selected physical error 
reference~\cite{yoder2017surface,azad2022surface}.

For either definition
$y\in\{\mathrm{meanT},\mathrm{Pmean}\}$, let
$t_{\mathrm{PT}}^{(y,d)}$ denote the crossing time satisfying
\begin{equation}
P_L^{(d)}
\left(
t_{\mathrm{PT}}^{(y,d)}
\right)
=
p_a
\left(
t_{\mathrm{PT}}^{(y,d)}
\right).
\label{eq:ppt_general}
\end{equation}
The corresponding probability pseudothreshold is
\begin{equation}
p_{\mathrm{PT}}^{(y,d)}
=
p_a
\left(
t_{\mathrm{PT}}^{(y,d)}
\right).
\label{eq:ppt_dual}
\end{equation}
Thus, the two quantities reported in this work are
$p_{\mathrm{PT}}^{(\mathrm{meanT},d)}$ and
$p_{\mathrm{PT}}^{(\mathrm{Pmean},d)}$.

Since the two physical error references are generally different,
their crossing times can also differ:
\begin{equation}
t_{\mathrm{PT}}^{(\mathrm{meanT},d)}
\neq
t_{\mathrm{PT}}^{(\mathrm{Pmean},d)}.
\end{equation}
The two pseudothresholds therefore represent distinct finite-distance references 
rather than two alternative labels for the same crossing point.
 Reporting both allows us to distinguish performance changes
associated with the average hardware decoherence scale from those
associated specifically with the spatial heterogeneity of the
physical error probabilities.

\section{XYZ Planar Codes}
\label{subsec:xyz_codes}

The XYZ planar codes considered in this work are obtained from the
parent planar code by a local Clifford deformation~\cite{dua2024clifford}. 
Let $M$ denote a selected set of data
qubits and define
\begin{equation}
U_M
=
\prod_{q\in M} C_{X,q},
\qquad
C_X=e^{+i\pi X/4}.
\label{eq:xyz_local_clifford}
\end{equation}
The single-qubit transformation satisfies
\begin{equation}
C_X X C_X^\dagger=X,
\qquad
C_X Z C_X^\dagger=Y.
\label{eq:xyz_pauli_transform}
\end{equation}
Hence the $X$-type stabilizers of the parent code remain unchanged,
whereas every $Z$ acting on a selected data qubit is replaced by
$Y$.

For a Pauli operator \(P\), \(\operatorname{supp}(P)\) denotes the set of data qubits on which \(P\) acts nontrivially.
The set $M$ is chosen so that each parent $Z$-type stabilizer
contains exactly one selected data qubit,
\begin{equation}
\left|
\supp(S_z^Z)\cap M
\right|
=1
\qquad
\text{for every }S_z^Z .
\label{eq:xyz_exact_cover}
\end{equation}
If $q_M(z)$ denotes this unique qubit, the corresponding transformed
generator is
\begin{equation}
S_z^{ZY}(M)
=
Y_{q_M(z)}
\prod_{
q\in\supp(S_z^Z)\setminus\{q_M(z)\}
}
Z_q .
\label{eq:xyz_zy_generator}
\end{equation}
Thus a weight-four bulk $Z$ stabilizer becomes $Z^3Y$, while a
weight-three boundary stabilizer becomes $Z^2Y$. The resulting
stabilizer group is simply
\begin{equation}
\mathcal S(M)
=
U_M\mathcal S_{\rm P}U_M^\dagger ,
\label{eq:xyz_stabilizer_group}
\end{equation}
where $\mathcal S_{\rm P}$ is the stabilizer group of the parent
planar code. Because unitary conjugation preserves commutation and
linear independence, this construction defines a valid stabilizer
code.

In the binary symplectic representation~\cite{gottesman1997stabilizer}, let $H_X$ and $H_Z$
denote the support matrices of the parent $X$- and $Z$-type
generators, respectively, and let
$\bm m\in\mathbb F_2^n$ be the indicator vector of $M$. With
\[
D_M=\operatorname{diag}(\bm m),
\]
the stabilizer matrix becomes
\begin{equation}
H(M)
=
\begin{pmatrix}
H_X & 0\\
H_ZD_M & H_Z
\end{pmatrix}.
\label{eq:xyz_check_matrix}
\end{equation}
The nonzero lower-left block represents the $X$ component of the
$Y=(1|1)$ operators introduced into the $ZY$ stabilizers. The XYZ
planar code is therefore non-CSS~\cite{padmanabhan2021noncss}.
Since the deformation is a tensor product of single-qubit Clifford
transformations, it also preserves the support and weight of every
Pauli operator. Consequently, the number of encoded logical qubits
and the code distance are unchanged, and every XYZ configuration
constructed in this way has parameters
\begin{equation}
[[2d^2-2d+1,1,d]].
\label{eq:xyz_code_parameters}
\end{equation}

The logical operators are obtained from the parent logical operators
by the same local Clifford transformation,
\begin{equation}
\overline L(M)
=
U_M
\overline L_{\rm P}
U_M^\dagger .
\label{eq:xyz_logical_transport}
\end{equation}
Because $C_X$ leaves $X$ unchanged, the logical-$X$ operator can
be chosen as
\begin{equation}
\overline X(M)
=
\overline X_{\rm P}
=
\prod_{q\in\gamma_X}X_q .
\label{eq:xyz_logical_x}
\end{equation}
For the parent logical-$Z$ path, each $Z$ is converted to $Y$
whenever the corresponding data qubit belongs to $M$, giving
\begin{equation}
\overline Z(M)
\doteq
\left(
\prod_{q\in\gamma_Z\setminus M} Z_q
\right)
\left(
\prod_{q\in\gamma_Z\cap M} Y_q
\right),
\label{eq:xyz_logical_z}
\end{equation}
where $\doteq$ denotes equality up to an irrelevant Pauli phase.
Thus the geometric support and weight of the parent logical operators
are preserved. Equivalently, if
\[
\bm\ell_{\overline X}^{({\rm P})}
=
(\bm\ell_X|0),
\qquad
\bm\ell_{\overline Z}^{({\rm P})}
=
(0|\bm\ell_Z),
\]
their binary symplectic representations after the deformation are
\begin{align}
\bm\ell_{\overline X}(M)
&=
(\bm\ell_X|0),
\label{eq:xyz_lx_vector}\\
\bm\ell_{\overline Z}(M)
&=
(
\bm m\odot\bm\ell_Z
\mid
\bm\ell_Z
),
\label{eq:xyz_lz_vector}
\end{align}
where $\odot$ denotes element-wise multiplication. Since these
operators are obtained by conjugating a valid parent logical pair,
they continue to commute with all stabilizers and anticommute with
each other.

For the two configurations used in the simulations, let the odd code
distance be
\begin{equation}
d=2m+1,
\qquad
m=\frac{d-1}{2}.
\end{equation}
To specify the spatial positions of the selected qubits, we label the
data-qubit lattice by
\[
A_{r,u},
\quad 0\le r,u\le d-1,
\qquad
B_{r,u},
\quad 0\le r,u\le d-2.
\]
The $A/B$ labels refer only to lattice positions and do not represent
different data qubit types.

For the original configuration, denoted XYZ-O, the selected set is
\begin{equation}
M_{\mathrm O}
=
\left\{
A_{r,2j+1}:
0\le r\le d-1,\;
0\le j\le m-1
\right\}.
\label{eq:xyz_original_mask}
\end{equation}
Thus the odd $A$-columns
$1,3,\ldots,d-2$ are selected in every row. Since every parent
$Z$-type check contains two horizontally adjacent $A$ qubits,
this pattern selects exactly one qubit from each such check.

The relocated configuration, denoted XYZ-R, retains the XYZ-O pattern
in the first $A$ row, shifts the selected $A$ sites by one lattice
position in the remaining rows, and uses alternating boundary $B$
qubits to complete the boundary checks:
\begin{align}
M_{\mathrm R}
={}&
\left\{
A_{0,2j+1}:
0\le j\le m-1
\right\}
\nonumber\\
&\cup
\left\{
A_{r,2j}:
1\le r\le d-1,\;
1\le j\le m-1
\right\}
\nonumber\\
&\cup
\left\{
B_{2j+1,u}:
0\le j\le m-1,\;
u\in\{0,d-2\}
\right\}.
\label{eq:xyz_relocated_mask}
\end{align}
This pattern also satisfies Eq.~(\ref{eq:xyz_exact_cover}) for every
$Z$-type stabilizer. Both configurations therefore contain the same
number of selected qubits,
\begin{equation}
|M_{\mathrm O}|
=
|M_{\mathrm R}|
=
\frac{d(d-1)}{2}.
\label{eq:xyz_y_number}
\end{equation}

Figure~\ref{fig:xyz_two_configs} illustrates the two configurations
for $d=5$. XYZ-O places the $Y$ components on alternating
$A$-columns throughout the lattice, whereas XYZ-R retains this
pattern only in the top row and relocates the remaining $Y$
components to the shifted interior $A$ sites and alternating
boundary $B$ sites. The $X$-check layer is identical in the two
codes and is omitted from the figure. Consequently, XYZ-O and XYZ-R
have the same lattice, stabilizer supports and weights, number of
$Y$ components, logical dimension, and code distance; their only
difference is the spatial placement of the $Y$ operators within the
$ZY$ stabilizers.

\begin{figure*}[t]
\centering
\begin{tikzpicture}[font=\small]

\newcommand{\DrawXYZBase}{%
  \foreach \r in {0,...,4}{%
    \foreach \c in {0,...,4}{%
      \coordinate (A-\r-\c)
      at ({1.05*\c},{-1.05*\r});
    }%
  }%

  \foreach \r in {0,...,3}{%
    \foreach \c in {0,...,3}{%
      \coordinate (B-\r-\c)
      at ({1.05*\c+0.525},{-1.05*\r-0.525});
    }%
  }%

  \foreach \r in {0,...,4}{%
    \foreach \c in {0,...,3}{%
      \coordinate (C-\r-\c)
      at ({1.05*\c+0.525},{-1.05*\r});

      \draw[gray!65,line width=0.45pt]
      (C-\r-\c)--(A-\r-\c);

      \pgfmathtruncatemacro{\cp}{\c+1}
      \draw[gray!65,line width=0.45pt]
      (C-\r-\c)--(A-\r-\cp);

      \ifnum\r>0
        \pgfmathtruncatemacro{\rm}{\r-1}
        \draw[gray!65,line width=0.45pt]
        (C-\r-\c)--(B-\rm-\c);
      \fi

      \ifnum\r<4
        \draw[gray!65,line width=0.45pt]
        (C-\r-\c)--(B-\r-\c);
      \fi
    }%
  }%
}

\newcommand{\DrawXYZNodes}{%
  \foreach \r in {0,...,4}{%
    \foreach \c in {0,...,4}{%
      \node[
        circle,
        draw,
        fill=white,
        inner sep=1.5pt
      ] at (A-\r-\c) {};
    }%
  }%

  \foreach \r in {0,...,3}{%
    \foreach \c in {0,...,3}{%
      \node[
        rectangle,
        draw,
        fill=white,
        inner sep=1.4pt
      ] at (B-\r-\c) {};
    }%
  }%

  \foreach \r in {0,...,4}{%
    \foreach \c in {0,...,3}{%
      \node[
        diamond,
        draw,
        fill=white,
        inner sep=1.8pt
      ] at (C-\r-\c) {};
    }%
  }%
}

\begin{scope}[xshift=0cm]

  \DrawXYZBase

  \foreach \r in {0,...,4}{%
    \foreach \c in {1,3}{%

      \pgfmathtruncatemacro{\cl}{\c-1}

      \draw[
        dashed,
        line width=1.0pt
      ]
      (A-\r-\c)--(C-\r-\cl);

      \draw[
        dashed,
        line width=1.0pt
      ]
      (A-\r-\c)--(C-\r-\c);
    }%
  }%

  \DrawXYZNodes

  \foreach \r in {0,...,4}{%
    \foreach \c in {1,3}{%
      \node[
        circle,
        draw,
        fill=black,
        inner sep=1.5pt
      ] at (A-\r-\c) {};
    }%
  }%

  \node[font=\bfseries]
  at (2.1,0.65)
  {(a) XYZ-O};

\end{scope}

\begin{scope}[xshift=6.4cm]

  \DrawXYZBase

  \foreach \c in {1,3}{%

    \pgfmathtruncatemacro{\cl}{\c-1}

    \draw[
      dashed,
      line width=1.0pt
    ]
    (A-0-\c)--(C-0-\cl);

    \draw[
      dashed,
      line width=1.0pt
    ]
    (A-0-\c)--(C-0-\c);
  }%

  \foreach \r in {1,...,4}{%

    \draw[
      dashed,
      line width=1.0pt
    ]
    (A-\r-2)--(C-\r-1);

    \draw[
      dashed,
      line width=1.0pt
    ]
    (A-\r-2)--(C-\r-2);
  }%

  \foreach \r in {1,3}{%

    \pgfmathtruncatemacro{\rp}{\r+1}

    \foreach \c in {0,3}{%

      \draw[
        dashed,
        line width=1.0pt
      ]
      (B-\r-\c)--(C-\r-\c);

      \draw[
        dashed,
        line width=1.0pt
      ]
      (B-\r-\c)--(C-\rp-\c);
    }%
  }%

  \DrawXYZNodes

  \foreach \c in {1,3}{%
    \node[
      circle,
      draw,
      fill=black,
      inner sep=1.5pt
    ] at (A-0-\c) {};
  }%

  \foreach \r in {1,...,4}{%
    \node[
      circle,
      draw,
      fill=black,
      inner sep=1.5pt
    ] at (A-\r-2) {};
  }%

  \foreach \r in {1,3}{%
    \foreach \c in {0,3}{%
      \node[
        rectangle,
        draw,
        fill=black,
        inner sep=1.4pt
      ] at (B-\r-\c) {};
    }%
  }%

  \node[font=\bfseries]
  at (2.1,0.65)
  {(b) XYZ-R};

\end{scope}

\begin{scope}[yshift=-5.4cm,xshift=1.0cm]

  \node[
    circle,
    draw,
    fill=white,
    inner sep=1.5pt
  ] (la) at (0,0) {};
  \node[right=2pt of la]
  {$A$ data qubit};

  \node[
    rectangle,
    draw,
    fill=white,
    inner sep=1.4pt
  ] (lb) at (2.5,0) {};
  \node[right=2pt of lb]
  {$B$ data qubit};

  \node[
    diamond,
    draw,
    fill=white,
    inner sep=1.8pt
  ] (lc) at (5.0,0) {};
  \node[right=2pt of lc]
  {$ZY$ check};

  \draw[
    gray!65,
    line width=0.45pt
  ]
  (7.0,0)--(7.7,0);
  \node[right] at (7.7,0)
  {$Z$ action};

  \draw[
    dashed,
    line width=1pt
  ]
  (9.2,0)--(9.9,0);
  \node[right] at (9.9,0)
  {$Y$ action};

\end{scope}

\end{tikzpicture}

\caption{
The two distance-$5$ XYZ planar code configurations used in the
simulations.
The $X$-check layer is identical in XYZ-O and XYZ-R and is omitted
for clarity.
Solid gray edges denote the $Z$ action of a $ZY$ stabilizer, whereas
dashed black edges denote its unique $Y$ action.
Filled markers identify the selected $Y$-support data qubits.
The circle and square data-qubit markers indicate only the $A/B$
indexing sublattices introduced for the general-$d$ construction and
do not represent different data qubit types.
}
\label{fig:xyz_two_configs}

\end{figure*}

\section{POSTERIOR MWPM DECODER}
\label{sec:pmwpm}

Decoding the XYZ code can be formulated as two successive MWPM problems associated with
the two Pauli error components. In a conventional MWPM treatment, these components are
decoded using their prior error probabilities, without exploiting the correlations revealed
by the measured syndrome.

The posterior minimum-weight perfect matching (pMWPM) decoder exploits this information
by using the syndrome obtained in the first decoding stage to update the error probabilities
used in the second stage. In particular, the measured \(ZY\)-syndrome is used to update the
probability that an error on a data qubit at a \(Y\)-support position contains a \(Z\)
component.

Let

\[
H_{ZY}=(H_Y|H_Z)
\]
denote the binary symplectic representation of the \(ZY\) stabilizer
checks. The matrix \(H_Z\) describes the \(Z\)-support of the
stabilizers, whereas \(H_Y\) identifies the data qubits on which the
stabilizers act as \(Y\).

A general independent Pauli channel is considered. For data qubit
\(j\), let

\[
p_{I,j},p_{X,j},p_{Y,j},p_{Z,j}
\]
be the corresponding Pauli probabilities satisfying
\[
p_{I,j}+p_{X,j}+p_{Y,j}+p_{Z,j}=1 .
\]

The biased IID channel used in the simulations is obtained as the
special case

\[
p_{X,j}=p_{Y,j}
=\frac{p}{2(\eta+1)},
\qquad
p_{Z,j}
=\frac{\eta p}{\eta+1}.
\]

The same decoder also applies to INID noise by directly using the
qubit-dependent probabilities.

For a data qubit \(c\) belonging to the \(Y\)-support, define

\[
\mathcal C(c)=
\{\alpha:H_Y[\alpha,c]=1\}.
\]

For all configurations considered here,

\[
|\mathcal C(c)|=2 .
\]

Let the two neighboring \(ZY\) checks be \(a\) and \(b\), and denote
their syndrome values by

\[
s_a,s_b\in\{0,1\}.
\]

The peripheral supports excluding the central qubit \(c\) are

\[
N_a=\operatorname{supp}(H_Z[a,:])\setminus\{c\},
\]

\[
N_b=\operatorname{supp}(H_Z[b,:])\setminus\{c\}.
\]

A peripheral qubit flips the corresponding \(ZY\) measurement if its error contains an \(X\) or \(Y\) component, 
i.e., if the error contributes non-trivially to the measurement outcome.
Hence,

\[
q_j=p_{X,j}+p_{Y,j}.
\]

For a neighboring check \(\alpha\in\{a,b\}\), the probabilities that an even or 
odd number of peripheral qubits produce flips are

\[
E_\alpha=
\frac{
1+\prod_{j\in N_\alpha}(1-2q_j)
}{2},
\]
and
\[
O_\alpha=
\frac{
1-\prod_{j\in N_\alpha}(1-2q_j)
}{2}.
\]

Define

\[
A_\alpha(0)=E_\alpha,
\qquad
A_\alpha(1)=O_\alpha .
\]

At the central qubit \(c\), an \(I\) or \(Y\) error commutes with
both neighboring \(ZY\) checks, whereas an \(X\) or \(Z\) error flips
both checks. Therefore,

\[
L_0(s_a,s_b)
=
A_a(s_a)A_b(s_b),
\]
corresponds to the case where the central qubit contributes no syndrome flip to either neighboring check,
and

\[
L_1(s_a,s_b)
=
A_a(1-s_a)A_b(1-s_b)
\]
corresponds to the case where the central qubit contributes flips to both neighboring checks.

The posterior probability that an error occurring on a data qubit contains a \(Z\) component 
is obtained using Bayes' rule. 
Since a \(Z\) component is present for either a physical \(Z\) or \(Y\) error,

\begin{equation}
\label{eq:general_posterior}
P_c^{(Z)}
=
P(Z_c\cup Y_c|s_a,s_b)
=
\frac{
p_{Y,c}L_0(s_a,s_b)
+
p_{Z,c}L_1(s_a,s_b)
}{
(p_{I,c}+p_{Y,c})L_0(s_a,s_b)
+
(p_{X,c}+p_{Z,c})L_1(s_a,s_b)
}.
\end{equation}

The corresponding matching weight is

\begin{equation}
\label{eq:posterior_weight}
w_{c,\mathrm{post}}^{(Z)}
=
\ln
\frac{
1-P_c^{(Z)}
}{
P_c^{(Z)}
}.
\end{equation}

For data qubits outside the \(Y\)-support, no syndrome-conditioned
update is performed. Their ordinary MWPM weight is

\begin{equation}
\label{eq:z_prior_weight}
w_j^{(Z)}
=
\ln
\frac{
1-(p_{Z,j}+p_{Y,j})
}{
p_{Z,j}+p_{Y,j}
}.
\end{equation}

For \(Y\)-support qubits, Eq.~\eqref{eq:posterior_weight} replaces
Eq.~\eqref{eq:z_prior_weight}.

\subsection*{A. Homogeneous-noise reduction}

For homogeneous noise, all peripheral qubits have identical error
probability

\[
q=p_X+p_Y .
\]

If a \(ZY\) stabilizer contains \(v\) peripheral qubits, 
where \(v \in \{2,3\}\) for the present XYZ construction, the parity probabilities reduce to

\[
E_v
=
\frac{
1+(1-2q)^v
}{2},
\]
and
\[
O_v
=
\frac{
1-(1-2q)^v
}{2}.
\]

For unequal syndrome outcomes,

\[
P(Z\cup Y|0,1)
=
P(Z\cup Y|1,0)
=
p_Y+p_Z .
\]

For equal syndrome outcomes,

\[
P(Z\cup Y|0,0)
=
\frac{
p_ZO_v+p_YE_v
}{
(p_I+p_Y)E_v+(p_X+p_Z)O_v
},
\]
and
\[
P(Z\cup Y|1,1)
=
\frac{
p_ZE_v+p_YO_v
}{
(p_I+p_Y)O_v+(p_X+p_Z)E_v
}.
\]

These expressions recover the previously used homogeneous formulas
and show that the absence of reweighting for unequal syndrome pairs
is a consequence of the symmetric homogeneous limit rather than a
general property.

\subsection*{B. Sequential MWPM decoding}

Using the posterior weights above, the first decoding stage constructs
the weighted \(X\)-syndrome graph defined by \(H_X\):

\[
\hat{\bm e}_Z
=
\operatorname{MWPM}
(G_X(H_X,\bm w^{(Z)}),
\bm s_X).
\]

The estimated \(Z\)-component contributes to the measured \(ZY\)
syndrome through the \(Y\)-support of the stabilizers. Therefore,

\begin{equation}
\label{eq:zy_delta}
\bm\delta
=
H_Y\hat{\bm e}_Z
\pmod 2 ,
\end{equation}
and the residual syndrome is
\begin{equation}
\label{eq:residual_zy}
\bm s_{ZY}^{\prime}
=
\bm s_{ZY}\oplus\bm\delta .
\end{equation}

The second decoding stage estimates the \(X\)-component. The local
weight is

\begin{equation}
\label{eq:x_prior_weight}
w_j^{(X)}
=
\ln
\frac{
1-(p_{X,j}+p_{Y,j})
}{
p_{X,j}+p_{Y,j}
}.
\end{equation}

Using \(H_Z\), the residual syndrome, and these weights gives

\[
\hat{\bm e}_X
=
\operatorname{MWPM}
(G_{ZY}(H_Z,\bm w^{(X)}),
\bm s_{ZY}^{\prime}).
\]

\begin{algorithm}[t]
\caption{General posterior MWPM decoding}
\label{alg:general-pmwpm}

\KwIn{
\(\bm{s}_X,\bm{s}_{ZY}\);
\(H_X,H_{ZY}=(H_Y|H_Z)\);
local Pauli probabilities
\(\{p_{I,j},p_{X,j},p_{Y,j},p_{Z,j}\}\)
}

\KwOut{
\((\hat{\bm e}_Z,\hat{\bm e}_X)\)
}

Initialize ordinary \(Z\)-component weights.

For each \(Y\)-support qubit, compute
\(P_c^{(Z)}\) using Eq.~\eqref{eq:general_posterior}
and update the corresponding weight.

Construct \(G_X\) and obtain

\[
\hat{\bm e}_Z
=
MWPM(G_X,\bm{s}_X).
\]

Update the \(ZY\) syndrome using Eq.~\eqref{eq:residual_zy}.

Construct \(G_{ZY}\) using \(w_j^{(X)}\) and obtain

\[
\hat{\bm e}_X
=
MWPM(G_{ZY},\bm{s}_{ZY}^{\prime}).
\]

\Return
\((\hat{\bm e}_Z,\hat{\bm e}_X)\).

\end{algorithm}

The posterior probabilities depend only on local stabilizer
structures and single-qubit error probabilities. Therefore, all
possible local weights can be precomputed before Monte Carlo
sampling. During decoding, the additional operations consist only of the selection of the 
posterior weight according to the measured syndrome and the updating of the residual syndrome.

Consequently, pMWPM preserves the asymptotic complexity of sequential
MWPM decoding. The dominant computational cost remains the two MWPM
procedures, while posterior reweighting introduces only a linear
local overhead for bounded-weight stabilizers.

Throughout the remainder of this work, $O$ and $R$ denote the original 
and reoriented XYZ planar code geometries, respectively. Accordingly, 
XO-pMWPM denotes the original XYZ geometry decoded using pMWPM, whereas XR-pMWPM denotes the reoriented 
geometry decoded using the same pMWPM strategy. XO-MWPM denotes the original XYZ 
geometry decoded using conventional MWPM, while retaining all components of the pMWPM 
decoding framework except the posterior reweighting procedure.
Since different 
quantum processors may exhibit different noise-bias tendencies, the stabilizer structure 
can be further adapted to the corresponding hardware noise. In particular, replacing the 
appropriate $X$-type stabilizers with $XY$-type stabilizers yields a customized XYZ surface 
code designed to better accommodate $X$-biased noise. Accordingly, for the processor-dependent 
studies presented below, processor-specific XYZ planar code constructions tailored to the bias 
characteristics of the corresponding hardware are employed.

\section{XZZX MWPM DECODING}

As a baseline for comparison with the XYZ results, the degeneracy-aware matching decoder 
for the XZZX code introduced in Ref.~\cite{bonilla2021xzzx} is used. 
This also allows assessment of the impact of spatial reliability information on XZZX decoding.
The XZZX syndrome is decoded through primal and dual MWPM
components, with the assignment of the two directional error rates reversed between them.

For homogeneous biased Pauli noise, let
\begin{equation}
p=p_X+p_Y+p_Z,
\end{equation}
and define
\begin{equation}
p_{\mathrm{hr}}=p_Y+p_Z,
\qquad
p_{\mathrm{lr}}=p_X+p_Y,
\end{equation}
with corresponding step weights
\begin{equation}
w_{\mathrm{hr}}=\log\frac{1-p}{p_{\mathrm{hr}}},
\qquad
w_{\mathrm{lr}}=\log\frac{1-p}{p_{\mathrm{lr}}}.
\end{equation}
The primal component assigns the low- and high-rate probabilities to the horizontal and
complementary lattice orientations, respectively, while the dual component uses the
opposite assignment. For two syndrome vertices separated by \(n_{\mathrm{hr}}\) high-rate
and \(n_{\mathrm{lr}}\) low-rate steps, the degeneracy-aware pair weight is
\begin{equation}
W=
n_{\mathrm{hr}}w_{\mathrm{hr}}
+n_{\mathrm{lr}}w_{\mathrm{lr}}
-\log\binom{n_{\mathrm{hr}}+n_{\mathrm{lr}}}{n_{\mathrm{hr}}},
\end{equation}
where the last term accounts for equivalent shortest paths. This homogeneous decoder is
denoted XZZX-MWPM.

To study XZZX decoding under INID noise, two schemes are considered, 
both using the same code geometry and MWPM framework but differing in the spatial reliability information 
supplied to the decoder. The first, XZZX-A, uses the spatially averaged Pauli probabilities
\begin{equation}
\overline{p}_{\alpha}
=
\frac{1}{n}\sum_{i=1}^{n}p_{\alpha,i},
\qquad
\alpha\in\{X,Y,Z\},
\end{equation}
to construct the homogeneous XZZX matching weights. The physical errors are nevertheless
sampled from the full INID channel. Thus, XZZX-A provides the decoder only with averaged
error information while retaining the heterogeneous physical noise.

The second scheme, XZZX-I, uses qubit-specific reliability information. For data qubit
\(i\), define
\begin{equation}
p_i=p_{X,i}+p_{Y,i}+p_{Z,i},
\end{equation}
and
\begin{equation}
p_{\mathrm{hr},i}=p_{Y,i}+p_{Z,i},
\qquad
p_{\mathrm{lr},i}=p_{X,i}+p_{Y,i}.
\end{equation}
The same orientation rule as in the homogeneous decoder is applied locally: the primal
component assigns \(p_{\mathrm{lr},i}\) and \(p_{\mathrm{hr},i}\) to the horizontal and
complementary orientations, respectively, while the dual component reverses this
assignment. A graph step associated with qubit \(i\) is assigned the weight
\begin{equation}
w_i=\log\frac{1-p_i}{p_{\mathrm{comp},i}},
\end{equation}
where \(p_{\mathrm{comp},i}\) is the component probability appropriate to its orientation
and primal/dual component.

Pair weights are then obtained from minimum-weight paths on this location-dependent graph.
If \(d_{\min}(u,v)\) is the minimum path weight between syndrome vertices \(u\) and \(v\),
and \(N_{\min}(u,v)\) is the number of minimum-weight paths, the pair weight is
\begin{equation}
W(u,v)=d_{\min}(u,v)-\log N_{\min}(u,v).
\end{equation}
For spatially uniform error probabilities,
\(N_{\min}=\binom{n_{\mathrm{hr}}+n_{\mathrm{lr}}}{n_{\mathrm{hr}}}\), and this expression
reduces to the homogeneous degeneracy-aware weight above.

XZZX-A and XZZX-I therefore differ only in the spatial information used to construct the
matching weights: XZZX-A uses spatially averaged Pauli probabilities, whereas XZZX-I
retains qubit-specific probabilities. Their comparison isolates the effect of spatial
reliability information under the same INID physical channel.

For consistency, XO-MWPM supplied with qubit-specific likelihood weights under INID noise
is denoted XO-MWPM-I.

\begin{figure*}[t]
    \centering
    \includegraphics[width=1\textwidth]
    {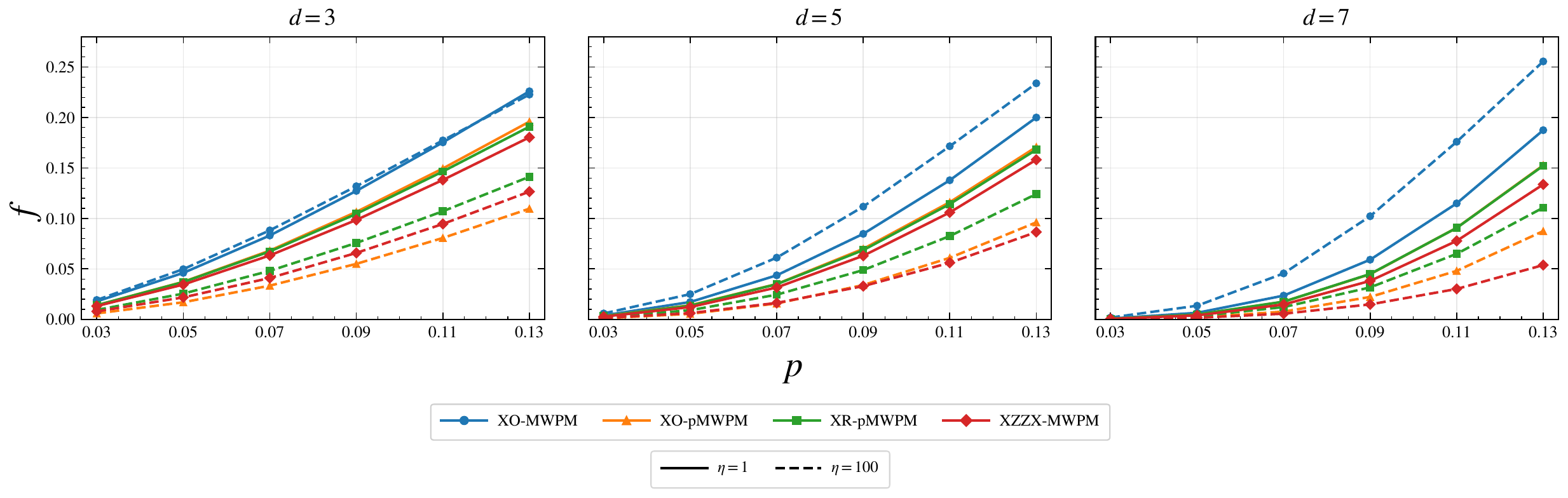}
    \caption{
    Logical failure rate $f$ as a function of the physical error rate $p$
    at code distances $d=3$, $5$, and $7$.
    Results are shown for XO-MWPM, XO-pMWPM, XR-pMWPM, and XZZX-MWPM
    under low-bias noise ($\eta=1$, solid lines) and strongly biased noise
    ($\eta=100$, dashed lines).
    The comparison illustrates the joint dependence of the relative
    performance on noise bias and code distance.
    }
    \label{fig:fixed-d-logical-error}
\end{figure*}

\section{RESULTS}
\subsection*{A. Logical error performance at fixed code distance}

The logical error performance of the different code-decoder combinations 
is compared at fixed code distance.
Figure~\ref{fig:fixed-d-logical-error} shows the logical failure rate
$f$ as a function of the physical error rate $p$ for
$d=3$, $5$, and $7$ under low-bias ($\eta=1$) and strongly biased
($\eta=100$) Pauli noise.

Under low-bias noise ($\eta=1$), XZZX-MWPM generally gives the lowest logical
failure rate over the distances and physical error rates considered.
The static XO-MWPM decoder performs substantially worse, whereas
posterior reweighting markedly improves the XYZ planar code:
both XO-pMWPM and XR-pMWPM reduce the logical failure rate relative
to XO-MWPM.

The ordering changes as the noise becomes strongly biased.
For $\eta=100$, XO-pMWPM becomes highly competitive with XZZX-MWPM
and can outperform it at small code distance. This advantage is,
however, distance dependent. At $d=5$ the two schemes become comparable
over part of the sampled $p$ range, whereas at $d=7$ the advantage of
XZZX is largely restored. The fixed-distance data therefore indicate
a bias- and distance-dependent crossover rather than a universal
ordering between the XYZ and XZZX constructions.

The intermediate bias values confirm that this crossover is systematic
rather than specific to $\eta=100$. At $d=3$, XO-pMWPM is consistently
inferior to XZZX at $\eta=1$, but already outperforms it at all
sampled physical error rates for $\eta=10$. At $\eta=1000$, its relative
advantage increases to approximately $15$--$25\%$. At $d=5$, the
XO-pMWPM advantage is mainly confined to the low-$p$ regime, whereas
at $d=7$ XZZX again dominates except near the lowest sampled error rate.
XR-pMWPM, although substantially superior to static XO-MWPM, 
does not surpass XZZX over the \(\eta=10\) and \(1000\) data considered here.

These results establish the first important trend: homogeneous IID noise
does not lead to a distance-independent decoder ranking. In particular,
strong bias creates a short-distance regime in which a bias-aware
XYZ decoder can outperform XZZX, motivating a more systematic comparison
of the decoding strategies and their pseudothreshold behavior.

\subsection*{B. Thresholds and pseudothresholds under IID noise}

To separate the contribution of posterior reweighting from that of the
orientation of the $Y$ operators, three decoding settings are first studied 
for the XYZ planar code under the homogeneous biased Pauli noise. XO-MWPM and
XO-pMWPM use the same original code geometry and differ only in the
decoding strategy, whereas XO-pMWPM and XR-pMWPM use the same pMWPM
decoder but different orientations of the $Y$ operators. These two
comparisons therefore provide controlled ablations of posterior
reweighting and geometric reorientation, respectively.

First, the corresponding asymptotic IID threshold is established using finite-size scaling.
Following the critical-exponent approach of
Ref.~\cite{wang2003confinement}, the scaling variable is taken as
\begin{equation}
    x=(p-p_c)d^{1/\nu},
\end{equation}
and the logical failure rate near the critical point is fitted to
\begin{equation}
    f=A+Bx+Cx^2,
\end{equation}
where $p_c$ and $\nu$ are fitted together with the coefficients
$A$, $B$, and $C$. The applicability of this finite-size-scaling
description near the critical region is discussed in
Ref.~\cite{watson2014surface}.

Because only finite code distances are computationally accessible,
the extracted $p_c$ values should be interpreted as finite-size-scaling
estimates of the asymptotic threshold. In particular, small code
distances can shift the apparent critical point~\cite{xiao2024exact},
and corrections to the asymptotic scaling form can remain relevant when
the simulated distances are not sufficiently large.

\begin{figure}[htbp]
    \centering
    \begin{minipage}[b]{0.48\textwidth}
        \centering
        \includegraphics[width=\linewidth]{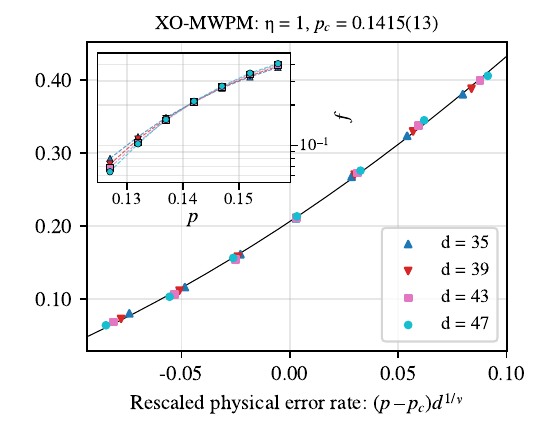}
    \end{minipage}
    \hfill
    \begin{minipage}[b]{0.48\textwidth}
        \centering
        \includegraphics[width=\linewidth]{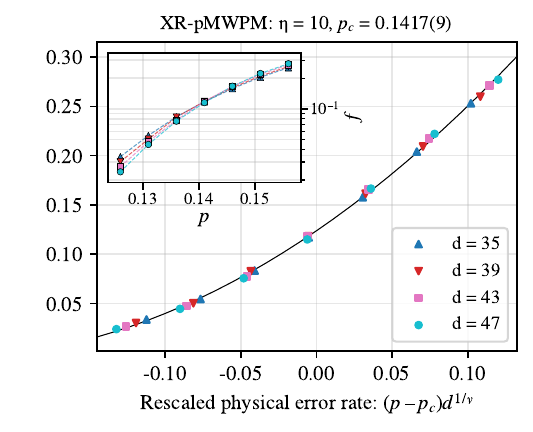}
    \end{minipage}

    \vskip\baselineskip

    \begin{minipage}[b]{0.48\textwidth}
        \centering
        \includegraphics[width=\linewidth]{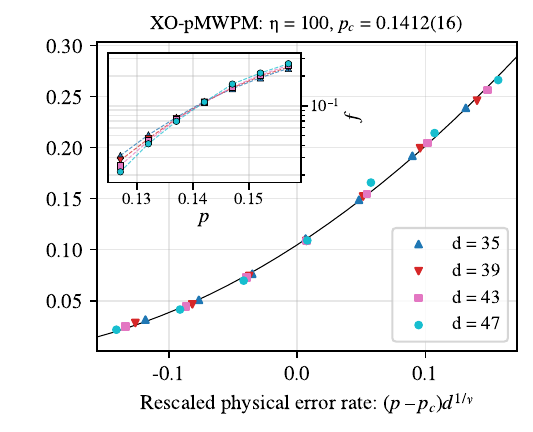}
    \end{minipage}
    \hfill
    \begin{minipage}[b]{0.48\textwidth}
        \centering
        \includegraphics[width=\linewidth]{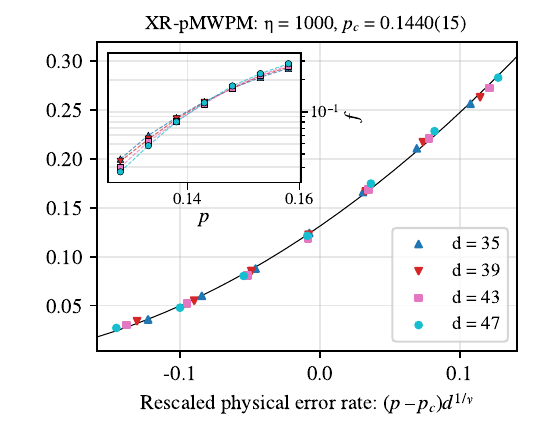}
    \end{minipage}

    \caption{
    Finite-size scaling of the logical failure rate for
    $\eta\in\{1,10,100,1000\}$.
    The data are plotted as a function of
    $x=(p-p_c)d^{1/\nu}$, and the solid curves show the best fits to
    $f=A+Bx+Cx^2$. Insets show the corresponding unscaled Monte Carlo
    results for different code distances.
    }
    \label{fig:scaling_all}
\end{figure}

The resulting threshold estimates are summarized in
Table~\ref{tab:compare_methods}. Replacing conventional MWPM by pMWPM
while keeping the XO geometry unchanged produces a substantial
improvement at every bias considered. The relative increase in $p_c$
is approximately $9.8\%$, $32.1\%$, $37.2\%$, and $38.7\%$ for
$\eta=1$, $10$, $100$, and $1000$, respectively. Thus, the benefit
of posterior reweighting becomes progressively more important as the
noise bias increases. Whereas the XO-MWPM threshold decreases from
approximately $14.2\%$ at $\eta=1$ to approximately $10.2\%$ at
$\eta=1000$, the XO-pMWPM threshold remains close to $14.2\%$
throughout the strongly biased regime.

In contrast, changing the orientation of the $Y$ operators while keeping
the pMWPM strategy fixed produces a smaller effect. At $\eta=1$ and
$\eta=10$, the XO-pMWPM and XR-pMWPM estimates agree within their
statistical uncertainties. At $\eta=100$ and $1000$, XR-pMWPM exceeds
XO-pMWPM by approximately $0.002$. The separation is less conclusive
at $\eta=100$ because of the comparatively large uncertainty of the
XO-pMWPM estimate, but is considerably clearer at $\eta=1000$.
Posterior reweighting therefore accounts for the larger of the two threshold changes examined here, 
while $Y$-operator reorientation provides a secondary,
bias-dependent improvement.

Representative finite-size-scaling results are shown in
Fig.~\ref{fig:scaling_all}. The crossings of the finite-distance curves
occur close to the fitted critical points, and the rescaled data show
good qualitative collapse in the critical region, supporting the use
of the quadratic approximation for extracting the threshold estimates.

Across the wider bias range considered in this work, 
the pMWPM-decoded XYZ planar code exhibits a pronounced high value in the low-bias 
regime and then remains within a relatively narrow threshold range of approximately 
14.1--14.4\% throughout the strongly biased regime.
The remaining
nonmonotonic bias dependence is consistent with competing effects:
increasing bias suppresses the $X$- and $Y$-error components that
perturb posterior reweighting, while simultaneously modifying the
structure and density of the relevant $Z$-check subgraphs.

\begin{table}[t]
    \centering
    \caption{
    Threshold estimates for XO-MWPM, XO-pMWPM, and XR-pMWPM
    under biased IID Pauli noise.
    }
    \label{tab:compare_methods}
    \begin{tabular}{c c c c}
        \toprule
        $\eta$ & XO-MWPM & XO-pMWPM & XR-pMWPM \\
        \midrule
        1     & 0.1415(13) & 0.1554(6)  & 0.1553(7) \\
        10    & 0.1074(5)  & 0.1419(5)  & 0.1417(9) \\
        100   & 0.1029(5)  & 0.1412(16) & 0.1432(4) \\
        1000  & 0.1024(3)  & 0.1420(3)  & 0.1440(15) \\
        \bottomrule
    \end{tabular}
\end{table}

However, the threshold alone cannot fully characterize the performance of 
experimentally relevant finite-distance codes. 
Therefore, the scaling analysis is complemented with the finite-distance pseudothreshold.
Throughout this section and the following section, the term pseudothreshold refers 
to the nominal pseudothreshold unless otherwise stated. The nominal pseudothreshold 
is extracted from the complete, unresampled simulation data using the 
finite-distance crossing condition, with the crossing location determined 
by linear interpolation between the nearest pair of adjacent sampled points with 
opposite signs that bracket the zero crossing.

\begin{figure*}[t]
    \centering
    \includegraphics[width=0.98\textwidth]
    {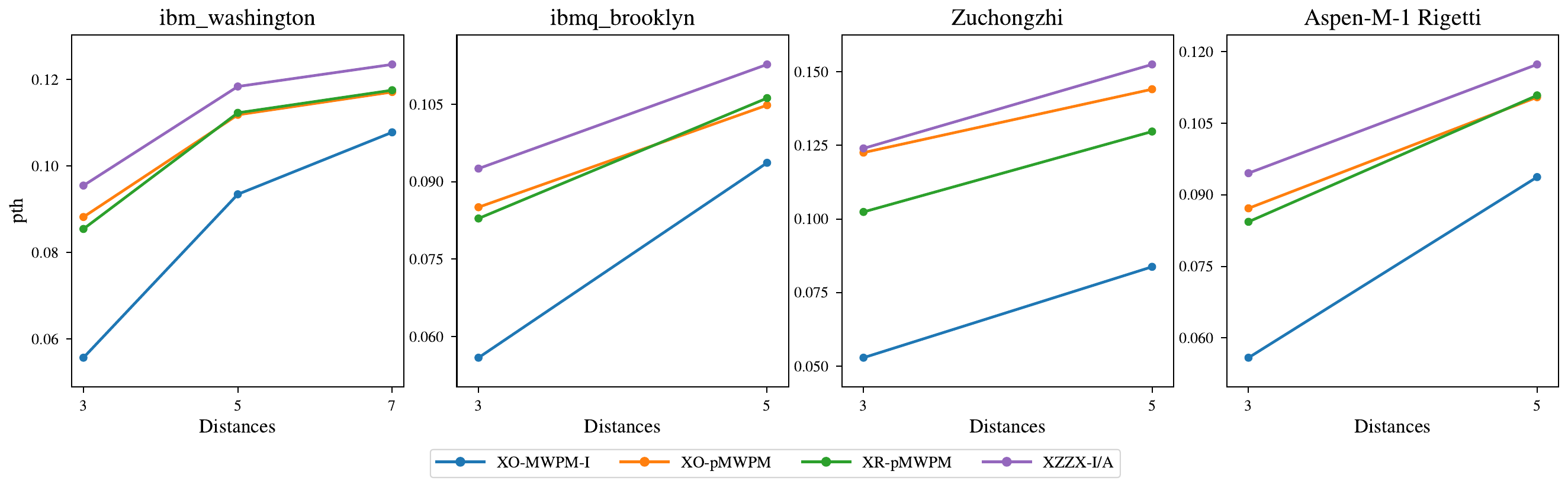}
    \caption{
    Finite-distance IID pseudothreshold as a function of code distance
    for homogeneous reference channels constructed using \(P\)-mean.
    Results are shown for IBM Washington, IBM Brooklyn, Zuchongzhi,
    and Rigetti Aspen-M-1. The curves correspond to XO-MWPM-I,
    XO-pMWPM, XR-pMWPM, and XZZX-I/A. Under IID noise, the average-rate
    and per-qubit XZZX weightings are equivalent and are therefore
    denoted collectively by XZZX-I/A. Data are available for
    $d=3,5,7$ on IBM Washington and for $d=3,5$ on the other
    processors. 
    Only nominal pseudothreshold estimates are shown.
    }
    \label{fig:iid-pmean-distance}
\end{figure*}

\begin{figure*}[t]
    \centering
    \includegraphics[width=0.98\textwidth]
    {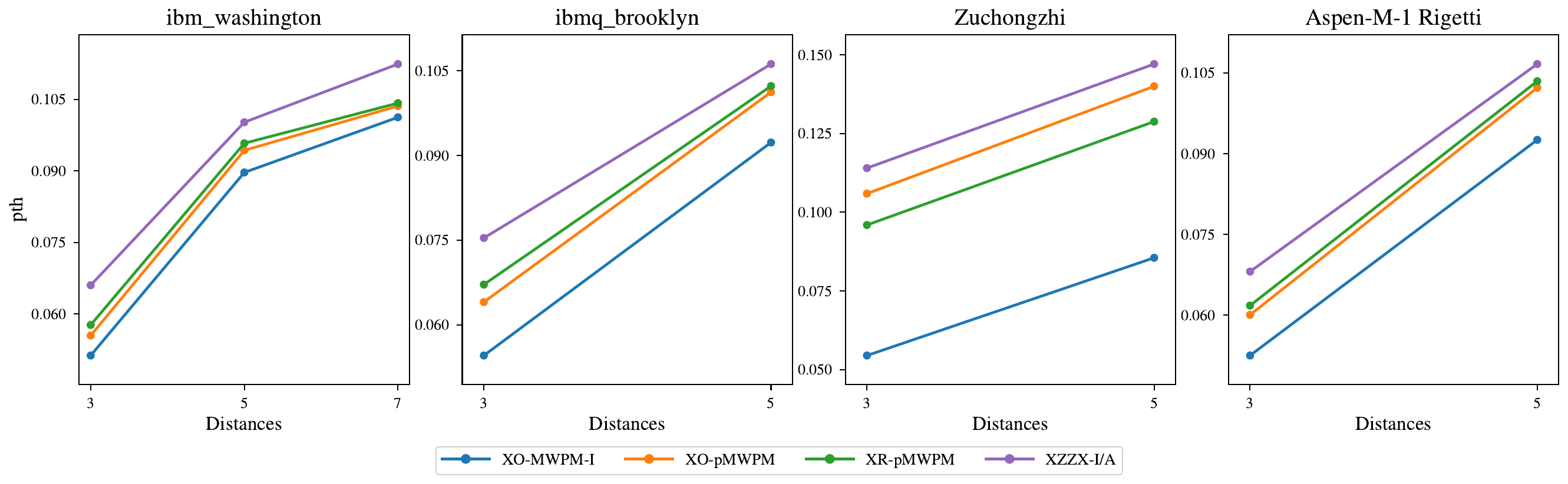}
    \caption{
    Finite-distance IID pseudothreshold as a function of code distance
    for homogeneous reference channels constructed using mean-$T$.
    The same processor set and code--decoder combinations as in
    Fig.~\ref{fig:iid-pmean-distance} are shown. 
    Only nominal pseudothreshold estimates are shown.
    }
    \label{fig:iid-meant-distance}
\end{figure*}

For the subsequent comparison with spatially non-identical hardware
noise, each processor calibration is mapped to two corresponding homogeneous
IID reference channels following the \(P\)-mean and mean-$T$ definitions
introduced above. The former preserves the average Pauli error vector,
whereas the latter preserves the average coherence parameters.
The corresponding finite-distance IID pseudothresholds are shown in
Figs.~\ref{fig:iid-pmean-distance} and
\ref{fig:iid-meant-distance}.

Several features are common to the two IID constructions. First, the
pseudothreshold increases with code distance for every method and
processor over the distances considered. Second, XZZX-I/A provides the
largest IID pseudothreshold for every processor--distance combination.
Together, these results characterize the finite-distance behavior under 
homogeneous noise and provide a reference point for assessing how spatial 
noise nonuniformity modifies the corresponding pseudothresholds in the INID setting.

Posterior reweighting again provides a clear improvement within the XYZ
family. XO-pMWPM consistently exceeds XO-MWPM-I, with the enhancement being
particularly pronounced for Zuchongzhi and at small code distance.
For example, under \(P\)-mean at $d=3$, the Zuchongzhi pseudothreshold
increases from approximately $0.053$ for XO-MWPM-I to $0.123$ for
XO-pMWPM, bringing the latter close to the XZZX-I/A value of $0.124$.
Thus, the importance of incorporating the channel statistics into the
matching weights is visible not only in the asymptotic thresholds but
also in the finite-distance regime.

The effect of the $Y$ operators is more complex; unlike posterior reweighting, 
it does not produce a uniform finite-distance advantage. On IBM Washington, 
IBM Brooklyn, and Rigetti Aspen-M-1, XO-pMWPM and XR-pMWPM typically differ only slightly, 
and their ordering can change with distance and with the chosen IID construction. 
Zuchongzhi shows a clearer distinction, with XO-pMWPM remaining above XR-pMWPM under both 
\(P\)-mean and mean-$T$. 
Therefore, the modest asymptotic threshold improvement associated 
with the reoriented geometry under strong bias should not be interpreted as a universal 
improvement in finite-distance performance.

This distinction illustrates the complementary roles of the two metrics.
The critical threshold is governed by the large-distance scaling near
the phase transition, whereas the pseudothreshold is determined by the
finite-distance crossing of the logical and physical error rates.
Consequently, a geometry with a slightly higher asymptotic threshold
need not have the highest pseudothreshold, or the lowest logical failure
rate, at the code distances accessible in practice.

Finally, the different absolute pseudothresholds obtained from \(P\)-mean
and mean-$T$ already indicate that a finite-distance pseudothreshold depends
on how the physical noise is mapped onto a scalar error coordinate.
Here these two constructions are used only to establish the appropriate
homogeneous IID references. 
Their physical distinction becomes essential once spatial non-identicality 
is introduced, which is examined next.

\subsection*{C. Decoder, hardware, and distance dependence under INID noise}
\begin{figure*}[t]
    \centering
    \includegraphics[width=1\textwidth]
    {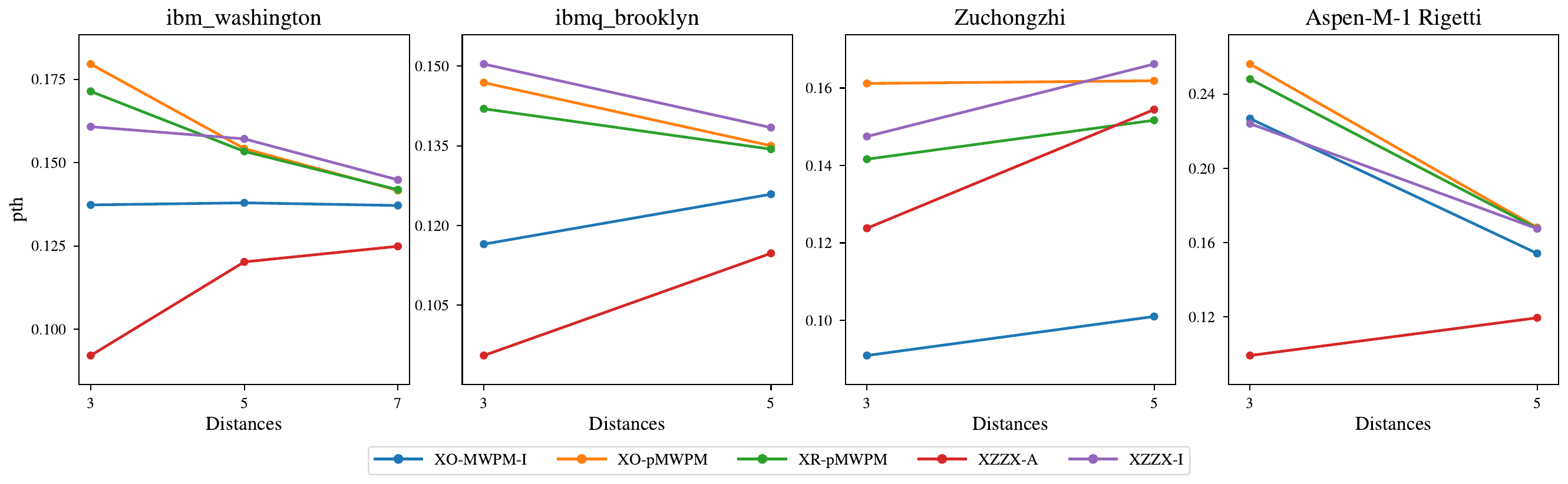}
    \caption{
    INID pseudothreshold as a function of code distance under the \(P\)-mean
    definition for IBM Washington, IBM Brooklyn, Zuchongzhi, and
    Rigetti Aspen-M-1.
    The curves correspond to XO-MWPM-I, XO-pMWPM, XR-pMWPM, XZZX-A, and XZZX-I.
    Only nominal pseudothreshold estimates are shown.
    }
    \label{fig:inid-pmean-distance}
\end{figure*}

\begin{figure*}[t]
    \centering
    \includegraphics[width=1\textwidth]
    {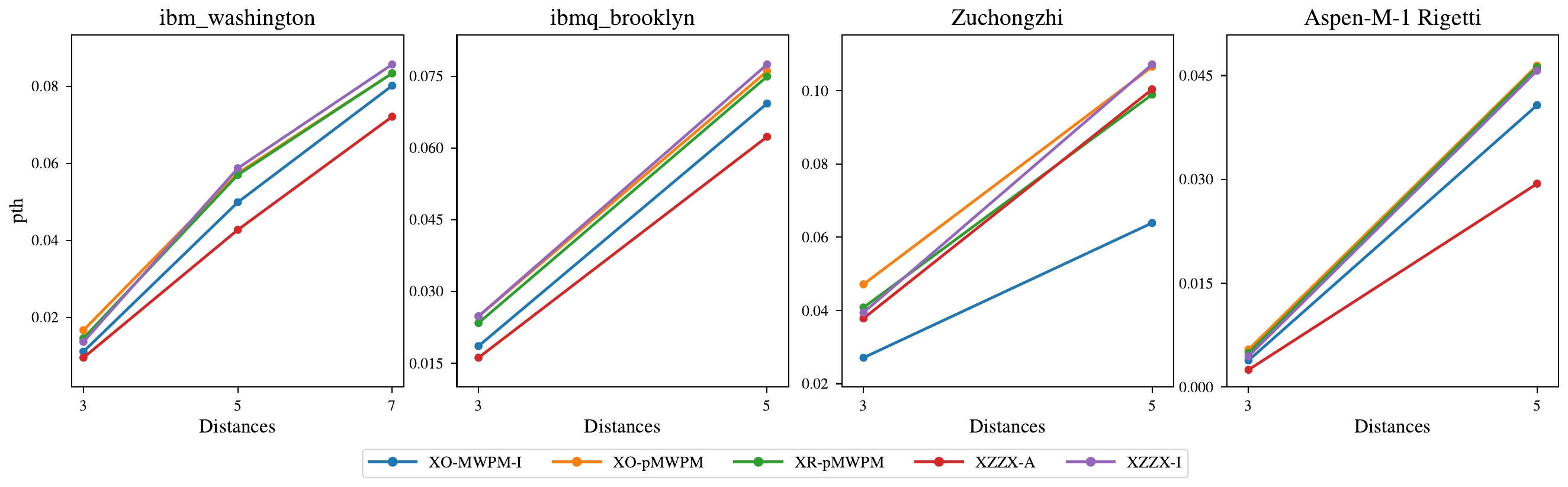}
    \caption{
    INID pseudothreshold as a function of code distance under the
    mean-$T$ definition for the four hardware platforms.
    The same code--decoder combinations as in
    Fig.~\ref{fig:inid-pmean-distance} are shown.
    Only nominal pseudothreshold estimates are shown.
    }
    \label{fig:inid-meant-distance}
\end{figure*}

Having introduced the two physical error scalarizations above, 
the focus turns to how spatially non-identical noise modifies the relative performance of 
the different code--decoder combinations and how this effect depends on hardware and code distance.
Figures~\ref{fig:inid-pmean-distance} and
\ref{fig:inid-meant-distance} show the INID pseudothresholds under the
\(P\)-mean and mean-$T$ definitions, respectively.

Under the corresponding homogeneous IID reference models, XZZX provides the largest pseudothreshold
for every processor--distance combination considered. Under INID noise,
however, this ordering is no longer universal. XO-pMWPM and XR-pMWPM become
substantially more competitive, and for several processor--distance
combinations they match or exceed XZZX-I. Thus, code--decoder schemes 
that are suboptimal under homogeneous noise can become competitive under 
spatially heterogeneous noise, particularly when the decoder incorporates 
qubit-specific reliability information.

The effect is particularly pronounced at small code distance.
Under \(P\)-mean at $d=3$, for example, XO-pMWPM reaches
$p_{\mathrm{PT}}=0.180$ and XR-pMWPM reaches $0.171$ on IBM Washington,
compared with $0.161$ for XZZX-I. On Rigetti Aspen-M-1, the corresponding
values are $0.256$, $0.248$, and $0.224$, respectively.
At $d=5$ on Rigetti, XO-pMWPM, XR-pMWPM, and XZZX-I become nearly
indistinguishable, with pseudothresholds of approximately $0.168$.
A similar relative trend is observed under mean-$T$, demonstrating that
the change in decoder ranking is not an artifact of a single
physical error scalarization.

The comparison between XZZX-A and XZZX-I more directly probes the role
of local reliability information because the code geometry is unchanged
while the decoding weights differ. 
Under INID noise, XZZX-I consistently outperforms XZZX-A, 
indicating that qubit-resolved weights provide a clear advantage 
over weights constructed from average error rates not only for planar codes~\cite{iolius2022performance} 
but also for the XZZX code.

Although the two scalarizations produce substantially different
absolute pseudothreshold values, the relative ordering of the decoding strategies
is considerably more stable. 
Among the nine processor--distance combinations
considered, the complete ordering of the five decoding strategies is unchanged
in six cases. In the remaining three cases, the difference is limited to a single
exchange between two adjacent strategies : XO-pMWPM and XR-pMWPM at $d=7$ for IBM
Quantum Washington, XZZX-I and XR-pM at $d=3$ for Zuchongzhi, and XZZX-I
and XO-MWPM-I at $d=3$ for Rigetti Aspen-M-1. 
Notably, the highest-ranked strategy is unchanged for all
nine processor--distance combinations. Thus, although the scalarization used to
represent the heterogeneous physical noise strongly affects the numerical value
of the pseudothreshold, its effect on the comparative ordering of the decoding
strategies is much more limited.
Next, the origin of this scalarization dependence is examined by comparing each INID model 
directly with its corresponding IID reference.

\subsection*{D. Scalarization dependence of the IID--INID pseudothreshold comparison}

\begin{figure*}[t]
    \centering
    \includegraphics[width=1\textwidth]
    {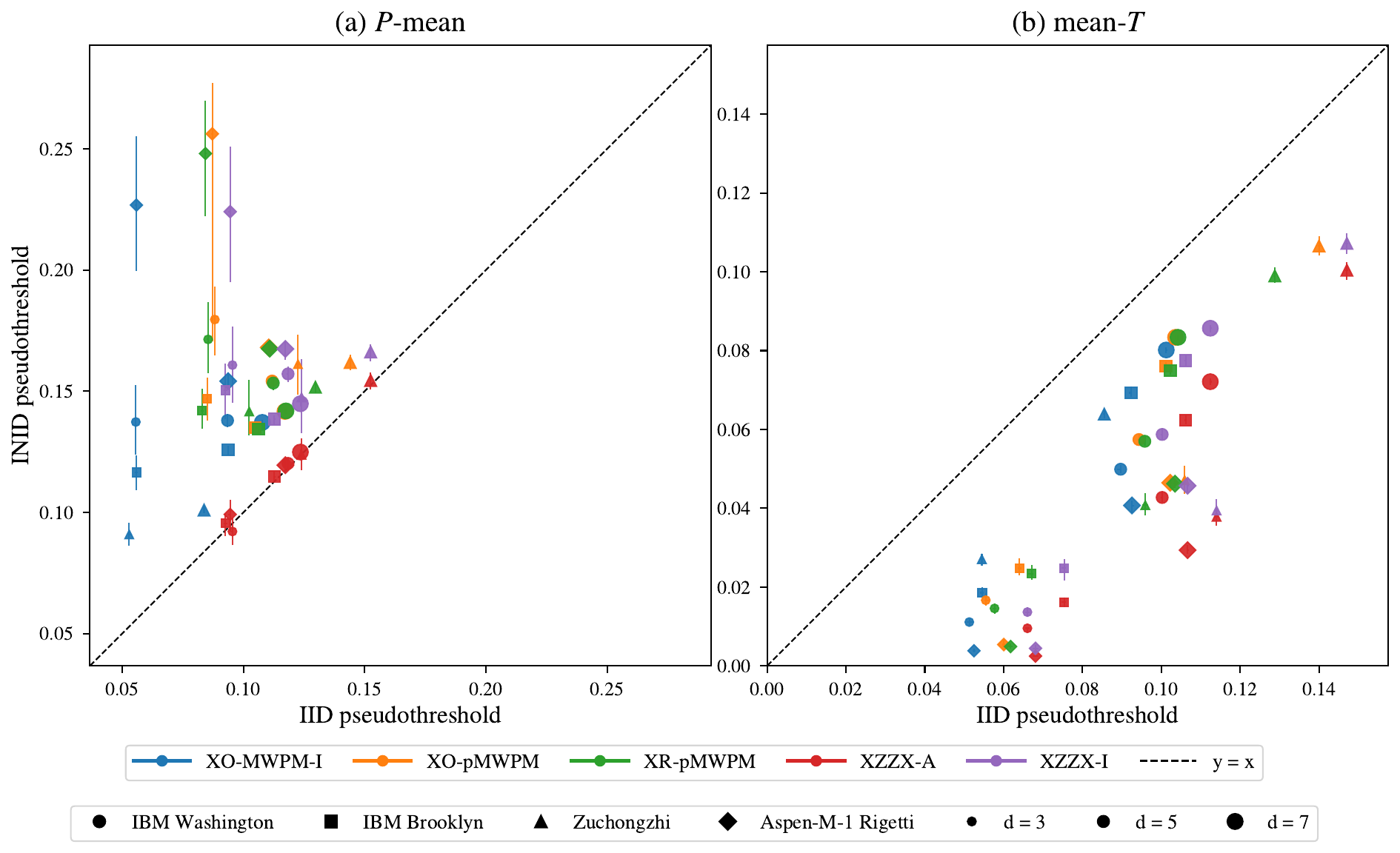}
    \caption{
    Comparison of IID and INID pseudothresholds under the two
    physical error definitions:
    (a) \(P\)-mean, $p_{\mathrm{P}}(t)=\langle p_i(t)\rangle$, and
    (b) mean-$T$,
    $p_{\mathrm{T}}(t)=p(\langle T_1\rangle,\langle T_2\rangle;t)$.
    Colors distinguish code--decoder combinations, marker shapes denote
    processors, and marker sizes denote code distances.   
    Horizontal and vertical error bars denote 95\% bootstrap confidence intervals 
    for the pseudothresholds on the corresponding axes, obtained by Monte Carlo 
    block resampling for IID noise and hierarchical resampling over spatial configurations 
    and logical error realizations for INID noise. Markers indicate the corresponding 
    nominal estimates.
    }
    \label{fig:iid-inid-comparison}
\end{figure*}

Figure~\ref{fig:iid-inid-comparison} summarizes the resulting comparison.
A pronounced difference in the predominant IID--INID trend appears between the two scalarizations. 
Under \(P\)-mean, 
the nominal INID pseudothreshold exceeds its
channel-matched IID value in 43 of the 45 comparisons considered.
Under mean-$T$, the nominal INID value lies below its IID reference
in all 45 cases.

This transformation arises because \(P\)-mean and mean-\(T\) define different physical error axes. 
Under \(P\)-mean, the IID reference preserves the mean
Pauli error vector of the heterogeneous device. 
Differences between the
IID and INID results therefore cannot be attributed simply to a change
in the mean physical error burden. 
Instead, they reflect the spatial heterogeneity of the error rates and the 
response of each decoder to that heterogeneity.
For decoders using qubit-specific error information,
this spatial structure can be incorporated through nonuniform decoding weights.

In contrast, the mean-\(T\) definition constructs a homogeneous reference by 
averaging coherence times before applying the nonlinear error model. For the hardware 
distributions considered here, \(p_{\mathrm T}(t)<p_{\mathrm P}(t)\). Thus, at the same elapsed 
time, the homogeneous mean-\(T\) reference carries a smaller average Pauli error burden than the 
actual heterogeneous channel. The IID--INID comparison is therefore not performed at fixed average 
Pauli error strength. This asymmetry helps account for the systematic downward displacement of the 
INID pseudothresholds relative to their IID references, as reflected by the points below the identity line.

Under the \(P\)-mean definition, XZZX-A provides a particularly informative empirical baseline. 
Across the nine available processor--distance combinations, the INID-to-IID pseudothreshold 
ratio has a mean of 1.014 and a median of 1.015, with individual relative changes ranging from 
approximately \(-3.5\%\) to \(+4.9\%\). The mean absolute IID--INID difference is \(2.3\times10^{-3}\). 
Thus, the XZZX-A points remain closely distributed around the identity line, indicating little systematic 
change in the pseudothreshold between the
\(P\)-mean IID reference and the corresponding INID case. Because XZZX-A uses averaged 
rather than qubit-specific error information in its decoding weights, its near-identity behavior provides a 
useful empirical baseline. For this scheme and the hardware instances considered, introducing spatial 
heterogeneity while preserving the mean Pauli error vector produces little systematic change in the pseudothreshold.

In contrast to XZZX-A, XZZX-I, XO-pMWPM, and XR-pMWPM show substantially larger upward displacements from 
the identity line under \(P\)-mean. 
The larger upward displacement of XZZX-I is consistent with a benefit from incorporating local reliability information 
into the decoding weights. For XO-pMWPM and XR-pMWPM, no corresponding average-rate 
baseline is available. Their upward displacements therefore demonstrate a relative enhancement from IID to 
INID within each scheme, but do not by themselves isolate how much of that enhancement originates specifically 
from the use of qubit-specific information.

The relative INID--IID enhancement is strongest at small code distance for the reliability-aware schemes. 
Averaged over the four processors available at \(d=3\), the INID-to-IID ratios are 1.717 for XZZX-I, 2.005 
for XO-pMWPM, and 2.011 for XR-pMWPM. At \(d=5\), the corresponding averages decrease to 1.268, 1.328, and 1.328, 
respectively. 
The available data at \(d=7\), which are limited to IBM Washington, continue this trend, 
with ratios of 1.173, 1.209, and 1.207. In contrast, XZZX-A remains close to unity over the available distances. 
This contrast 
suggests that the large IID--INID differences at small distance are not simply a generic consequence of spatial 
heterogeneity. Rather, they are consistent with a greater opportunity for hardware-aware decoding to benefit from 
the particular spatial reliability structure of small code blocks, with this relative enhancement becoming less 
pronounced as the code distance increases. This interpretation is also consistent with the observation that XO-pMWPM 
and XR-pMWPM, although less competitive than XZZX under the corresponding IID references, approach or exceed XZZX-I 
for several processor--distance combinations under INID noise.

The statistical uncertainty of the INID pseudothresholds also shows a distance dependence, particularly under 
\(P\)-mean. 
Averaged over the available estimates, the mean \(95\%\) confidence-interval width decreases 
from approximately \(0.030\) at \(d=3\) to \(0.0052\) at \(d=5\) and to \(0.0021\) for the 
available data at \(d=7\), which are limited to IBM Washington.
Because the hierarchical bootstrap incorporates variation among hardware configurations 
in addition to Monte Carlo sampling uncertainty, the wider intervals at small distance are consistent with 
a stronger dependence of the inferred pseudothreshold on the particular heterogeneous-noise realization. 
Notably, \(d=3\) is also the regime in which the largest nominal INID--IID enhancements are observed for 
the reliability-aware schemes. The coexistence of larger enhancement and greater uncertainty suggests that 
the relative INID advantage at small distance may be more configuration dependent. Both the nominal enhancement 
and the associated pseudothreshold uncertainty become smaller at the larger distances considered.

Previous studies based on mean-\(T\) references reported lower pseudothresholds for INID hardware noise compared 
with IID models \cite{iolius2022performance,iolius2023performance}. Our analysis shows that such comparisons depend 
on the chosen physical error scalarization. Because the mapping from coherence times to Pauli error probabilities is 
nonlinear, the mean-\(T\) construction can effectively compare the heterogeneous device with an idealized homogeneous 
channel defined on a more favorable physical error axis. Therefore, the observed reduction cannot be attributed to 
spatial heterogeneity alone and is strongly influenced by the choice of IID reference and physical error scalarization.

For a given INID simulation, the underlying logical error curve \(P_L^{\mathrm{INID}}(t)\) is identical under the 
two scalarizations; only the physical error coordinate used to define the crossing changes. The resulting difference 
between the \(P\)-mean and mean-\(T\) pseudothresholds therefore reflects the scalar representation of the same 
heterogeneous-noise simulation rather than a change in its logical error data. 
This distinction should be separated from the IID--INID comparison itself, for which the logical error curves generally differ.

Taken together, these results separate two distinct aspects of spatially non-identical hardware noise. 
The relative performance of the decoding strategies can change as different decoders respond differently to 
the spatial reliability structure, particularly at small code distance. In contrast, the absolute IID--INID 
pseudothreshold shift depends strongly on the physical error scalarization. Thus, decoder reordering is associated 
with differing responses to spatial noise structure, whereas the apparent IID--INID pseudothreshold advantage cannot 
be interpreted independently of the reference used to quantify the physical error rate.

\section{SIMULATIONS}
All simulations of XYZ planar codes were performed using the PyMatching package~\cite{higgott2022pymatching}, 
while simulations of XZZX planar codes used qecsim. For both symmetric and biased Pauli channels, 
the logical error probability was obtained by applying \(10^5\) random errors to the code. 
For the INID results, 
100 different qubit distributions (arrangements on the lattice) were considered for each point, 
and 15 probability values were considered for each specific point, 
including those for which the difference between \(P_L\) and \(p\) is less than 0.005. 
For each distribution and each probability value, \(10^4\) Monte Carlo simulations were performed, 
and bootstrap resampling over the 100 distributions was used to estimate the variation between configurations.

\section{CONCLUSION}

XYZ and XZZX planar codes have been investigated under homogeneous IID Pauli noise and a 
hardware-derived independent nonidentically distributed (INID) noise model, with particular attention 
to posterior-informed decoding and qubit-specific reliability information. Under homogeneous biased Pauli noise, 
posterior reweighting substantially increases the estimated asymptotic threshold of the XYZ planar code. 
With the XO geometry fixed, replacing conventional MWPM with pMWPM increases the threshold by 
approximately \(9.8\%\) at \(\eta=1\) and by \(32.1\%-38.7\%\) for \(\eta=10-1000\). In contrast, 
reorienting the \(Y\) operators while retaining pMWPM produces a much smaller change, with XO-pMWPM 
and XR-pMWPM agreeing within uncertainty at \(\eta=1\) and \(10\) and differing by only about \(0.002\) 
in threshold at \(\eta=100\) and \(1000\). The benefit of posterior reweighting also persists at finite 
distance: for example, under the \(P\)-mean IID reference for Zuchongzhi at \(d=3\), the pseudothreshold 
increases from approximately \(0.053\) for XO-MWPM-I to \(0.123\) for XO-pMWPM, close to the XZZX-I/A 
value of \(0.124\). The finite-distance comparison with XZZX further reveals a separate bias- and 
distance-dependent behavior, with pMWPM-decoded XYZ planar codes outperforming XZZX in parts of the 
small-distance, strongly biased regime while this advantage becomes less pronounced at the larger distances considered.

The hardware-derived results show that the INID noise model can also alter the relative performance 
of the code--decoder combinations compared with their corresponding homogeneous IID references. 
XZZX provides the largest pseudothreshold among the schemes considered for every available processor--distance 
combination under the IID references, whereas under INID noise XO-pMWPM and XR-pMWPM can match or exceed XZZX-I 
in selected cases. At \(d=3\) under \(P\)-mean, for example, XO-pMWPM reaches \(0.180\) compared with \(0.161\) 
for XZZX-I on IBM Washington, and \(0.256\) compared with \(0.224\) on Rigetti Aspen-M-1. 
The within-code XZZX-I/XZZX-A comparison provides a more direct assessment of qubit-specific reliability 
information because the code geometry is unchanged while the decoding weights differ. XZZX-I consistently 
yields higher INID pseudothresholds than XZZX-A for the hardware instances considered, supporting a systematic 
benefit from incorporating qubit-specific reliability information into the decoding weights.

A second central result concerns the interpretation of IID--INID pseudothreshold shifts. Previous 
studies using coherence-time-averaged IID references reported lower pseudothresholds under spatially 
nonuniform noise~\cite{iolius2022performance,iolius2023performance}. Consistent with the direction reported 
in those studies, all 45 mean-\(T\) comparisons considered here give a nominal INID pseudothreshold below the 
corresponding IID reference. Under \(P\)-mean, whose IID reference instead preserves the mean Pauli error vector 
of the heterogeneous channel, 43 of 45 comparisons give a nominal INID pseudothreshold above the corresponding 
mean-Pauli-matched IID value. These different predominant trends do not imply contradictory underlying INID 
logical error behavior: the INID logical error data are unchanged, whereas the physical error coordinate and 
associated IID reference differ. The IID--INID pseudothreshold shift is therefore not a reference-independent 
measure of spatial noise nonuniformity and must be interpreted together with the scalarization and reference 
construction used to define it.

Previous correlation-aware MWPM decoders exploit correlations between error components through matching-graph 
reweighting~\cite{iolius2023performance,delfosse2014decoding,fowler2013optimal}, while BP-assisted approaches use 
BP-derived probabilities to modify matching weights~\cite{higgott2023improved}. In pMWPM, the mixed-\(ZY\) syndrome 
is instead used to construct local syndrome-conditioned posterior weights before matching, followed by 
residual-syndrome decoding, without the iterative reweighting required by recursive MWPM~\cite{iolius2023performance}.

Overall, the results identify two complementary decoder-related effects. Posterior-informed weighting provides a 
substantial and bias-dependent improvement for the XYZ planar code, while the within-code XZZX comparison supports an 
additional benefit from using qubit-specific rather than average-rate reliability information under INID noise. 
More broadly, hardware nonuniformity can alter the relative competitiveness of the code--decoder combinations, 
while its apparent effect on the pseudothreshold depends on the physical error representation and homogeneous 
reference. 
The available data suggest stronger relative IID--INID enhancement at small distance, but the larger uncertainties 
in this regime and the limited data at \(d=7\) do not support a general distance-scaling conclusion.
Extending these comparisons to larger distances, additional hardware instances, and circuit-level or correlated 
noise will be important for determining how these effects persist in more realistic fault-tolerant settings.

\section{ACKNOWLEDGMENTS}
We are grateful to the anonymous referees and the associate editor for useful comments and suggestions that 
improved the presentation and quality of this paper. The work was supported by the National Natural Science 
Foundation of China under Grant No. 12271137.

\end{document}